\documentclass[aps,prl,twocolumn,pacs]{revtex4}
\usepackage[T1]{fontenc}
\usepackage[latin9]{inputenc}
\usepackage{amsmath}
\usepackage{graphicx}
\usepackage{amssymb}

\begin{document}

\title{Deterministic Generation of  Entangled Photons in Superconducting Resonator Arrays}

\author{Yong Hu and Lin Tian}

\email{ltian@ucmerced.edu}

\affiliation{University of California, Merced, 5200 North Lake Road, Merced, CA
95343, USA}

\date{\today }

\begin{abstract}
We present a scheme for the deterministic generation of entangled photon pairs in a superconducting resonator array.  The resonators form a Jaynes-Cummings Lattice via the coupling to superconducting qubits and the Kerr-like nonlinearity arises due to the coupling.  We show that entangled photons can be generated on-demand by applying spectroscopic techniques and exploiting the nonlinearity and symmetry in the resonators. The scheme is robust against small parameter spreads due to fabrication errors. Our findings can be used as a key element for quantum information processing in superconducting quantum circuits.
\end{abstract}  
\maketitle

Recent progresses in superconducting quantum circuits \cite{SCircuit} have enabled intensive study of the quantum behavior of the microwave cavity modes in superconducting resonators \cite{CQED}. Strong coupling between the cavity modes and superconducting qubits has been experimentally demonstrated in a number of systems \cite{CQEDexp}.  Quantum optical effects such as single photon manipulation and nonlinear spectrum have been observed \cite{CQED_QO}. Given their ultra-high quality factors, the superconducting resonators can be a powerful platform for quantum information processing and quantum state engineering  involving microwave photons. Recently, the Kerr nonlinearity was explored for the parametric amplification and squeezing of the resonator modes \cite{GirvinPDC,DevoretNPhys2010}. The generation of twin photons at different frequencies was studied via the pumping of superconducting qubits \cite{MarquardtPDC}. Meanwhile, coupled cavity arrays (CCA) can also be realized in superconducting systems to study quantum many-body effects such as the Mott insulator-to-superfluid transition \cite{CCA}.

One essential element for quantum information processing using microwave photons is the entangled photon source that can generate entangled photon pairs in a deterministic way and to distribute the photons in the circuits \cite{NielsenBook}. The entangled photons can be used to implement quantum teleportation and quantum cryptography and to test quantum mechanical principles such as the Bell inequalities \cite{EPR,BellPaper}. Various systems and approaches were explored to generate entangled photons in the past, including spontaneous parametric down conversion (SPDC) in quantum optical systems \cite{YShih}, pumping of bi-excitons and cavity-polaritons in semiconductors \cite{Semiconductor}, and most recently, using the atom-photon interface in coupled cavities \cite{JCho}. In those approaches, entanglement generation is either a stochastic process or a direct switching from the atomic to cavity states.

In this work, we present a novel approach for the deterministic generation of spatially-entangled or energy-entangled photons in superconducting resonators connected in a ring geometry.  Microwave photons can tunnel between adjacent resonators via capacitive coupling.  Each resonator is coupled to a superconducting qubit to form the Jaynes-Cummings (JC) Lattice, which can give rise to the Kerr-like nonlinearity in the resonator modes \cite{CavityQED,Imamoglu1997}.  By combining the nonlinearity, circuit symmetry, and a spectroscopic technique, we show that entangled photons can be generated with high fidelity in the presence of resonator dissipation and qubit decoherence. The nonlinearity and the symmetry are exploited to prevent off-resonant transitions which would otherwise impair the entanglement generation. This idea can be further explored to generate novel quantum states involving multiple photons and can also be applied to semiconductor micro-cavities. 

\begin{figure}
\includegraphics[clip,width=5.5cm]{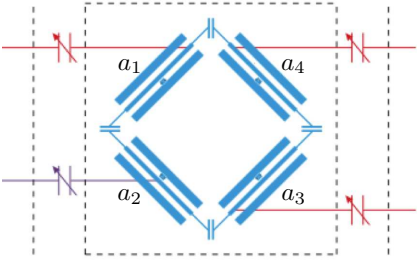} 
\caption{(Color online) Schematic circuit for the resonator array. Inside the dashed box: resonators where photons are generated; outside the box: interface to outside circuits.}
\label{fig1}
\end{figure}

The proposed circuit includes four superconducting resonators coupled in a ring geometry and each resonator couples to a superconducting qubit, as is shown in Fig.~\ref{fig1}. The system forms a JC Lattice with the total Hamiltonian $H_{t}=\sum_j ( H_{JC}^{(j)}+H_{c}^{(j)})$ which includes the JC model for each resonator-qubit system
\begin{equation}
H_{JC}^{(j)}=\hbar\omega_{0}a_{j}^{\dagger}a_{j}+(\hbar\omega_z/2)\sigma_{jz}+\hbar g (a_{j}^{\dagger}\sigma_{j-}+\sigma_{j+}a_{j})\label{eq:JC}
\end{equation}
and the tunneling between adjacent sites $H_{c}^{(j)}=\hbar J (a_{j}^{\dagger}a_{j+1}+a_{j+1}^{\dagger}a_{j}\ )$. Here, $a_{j}^{\dagger}$ ($a_{j}$) is the creation (annihilation) operator of the resonators, $\sigma_{jz},\,\sigma_{j\pm}$ are the Pauli matrices for the qubits, $\omega_{0}$  is the resonator frequency, $\hbar g$ is the coupling between the resonator and the qubit, and $\hbar J$ is the tunneling matrix element between the resonators. The eigenstates of the JC Model $H_{JC}^{(j)}$ are polariton states containing excitations in both the resonator and the qubit \cite{CQED}. With a finite coupling and a small detuning $\Delta=\omega_z-\omega_0$, the polariton spectrum demonstrates Kerr-like nonlinearity similar to the onsite interaction in the Bose-Hubbard Model. In the thermodynamic limit, the JC Lattice exhibits the Mott insulator-to-superfluid phase transition \cite{CCA, SachdevBook}. 

\begin{figure}
\includegraphics[clip,width=6.6cm]{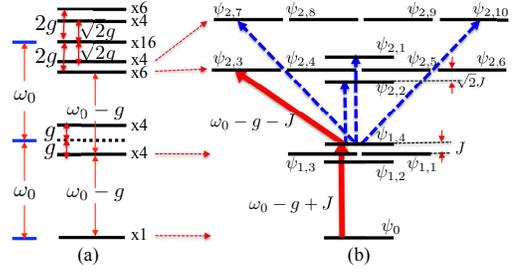} 
\caption{(Color online) (a) Eigenenergies at $J=\Delta=0$. The degeneracies are labelled to the right. (b) Selected eigenstates at finite $J$ and $\Delta=0$. Solid arrows are the designed transitions to generate $|\psi_{2,3}\rangle$ and dashed arrows are off-resonant transitions. }
\label{fig2}
\end{figure}

The nonlinearity and the symmetry in the JC Lattice  play an important role in the eigenenergy spectrum of the resonator array \cite{EPAPS}. The eigenstates can be labelled by the total number of excitations $N=\sum_j(a_{j}^{\dagger}a_{j}+\sigma_{j+}\sigma_{j-})$ which is a good quantum number with $[N,\, H_{t}]=0$. In Fig.~\ref{fig2}, the spectra for the low-lying eigenstates at the strong-coupling limit $g \gg J$ and $\Delta=0$ are shown. At $J=0$ without the tunneling, the eigenstates are product states of the polariton modes in a single JC Model \cite{EPAPS}. At finite (but weak) $J$, the tunneling Hamiltonian is treated as a perturbation. For $N=0$, we have the ground state $|\psi_{0}\rangle=\Pi_j |0\downarrow\rangle_{j}$. For $N=1$, the lowest four states $|\psi_{1,m}\rangle$ with $1\le m\le 4$ are superpositions of the four lower polariton states $L_{j,1}^{\dag}|\psi_{0}\rangle$. The tunneling Hamiltonian lifts the degeneracy of these states with energy splittings $\hbar J$. Other states for $N=1$ are the higher polariton modes $T_{j,1}^{\dag}|\psi_{0}\rangle$ which are separated from the lowest states by an energy $\sim \hbar g$. For $N=2$, the lowest six states $|\psi_{2,m}\rangle$ for $1\le m\le 6$ are superpositions of the states $L_{j,1}^{\dag}L_{k,1}^{\dag}|\psi_{0}\rangle$ with $j\ne k$ and contain two excitations occupying different sites. The tunneling Hamiltonian generates energy splittings of $\sqrt{2} \hbar J$ among these states. All other states for $N=2$ either involve excitations in the higher polariton modes or involve two excitations in the same site, such as $|\psi_{2,m}\rangle=L_{j,2}^{\dag}|\psi_0\rangle$ for $7\le m\le10$, with an energy $\sim\hbar  g$ above the lowest six states. 

Entangled photon pairs can be generated in the JC Lattice by spectroscopic techniques in the strong-coupling limit. The effective nonlinear interaction together with the tunneling $J$ lifts the degeneracy of the eigenstates and makes it possible to generate novel quantum states by applying microwave pulses with selected phases and frequencies. Assume a monochromatic driving $H_{d}=\hbar\sum_{j=1}^{4}(\epsilon_{j}e^{-i\omega_{d}t}a_{j}^{\dagger}+\epsilon_{j}^{\ast}e^{i\omega_{d}t}a_{j})$ being applied to the resonators with the amplitudes $\epsilon_{j}$ and frequency $\omega_{d}$. This pulse can generate transitions between eigenstates differing by one excitation number.  When $|\epsilon_{j}|\ll J, g$, only resonant transitions are important. Furthermore, we can choose the relative phases and amplitudes of $(\epsilon_{1},\epsilon_{2},\epsilon_{3},\epsilon_{4})$ to design the allowed transitions.  Consider the eigenstate at $\Delta=0$
\begin{equation}
|\psi_{2,3}\rangle=(L_{1,1}^{\dag}L_{3,1}^{\dag}-L_{2,1}^{\dag}L_{4,1}^{\dag})|\psi_{0}\rangle/\sqrt{2}\label{eq:psi23}
\end{equation}
which contains two excitations in the lower polariton modes. This state is an EPR state with spatially-entangled polaritons if we map the modes $L_{1,1}$ and $L_{4,1}$ to one polarization and the modes $L_{2,1}$ and $L_{3,1}$ to the orthogonal polarization \cite{YShih}. Below we construct a two-pulse process to generate this state with high fidelity.  Initially,  with $\hbar\omega_0 \gg k_BT$ in the cryogenic environment, the system is prepared to the ground state $|\psi_0\rangle$ by thermalization. The first pulse has uniform amplitudes $(\epsilon_{1}, \epsilon_{2}, \epsilon_{3}, \epsilon_{4})=\epsilon(1,1,1,1)$ with $\epsilon\ll J$ and the pumping  frequency $\omega_{d}=\omega_{0}-g+J$. This pulse induces a resonant transition and hence a Rabi oscillation between $|\psi_{0}\rangle$ and the single-polariton eigenstate $|\psi_{1,4}\rangle= (\sum L_{j,1}^\dag \psi_0)/2$, as is indicated by the solid arrow in Fig.~\ref{fig2}(b). After applying the pulse for a duration of $\pi/2\sqrt{2}\epsilon$, the state is then pumped to $|\psi_{1,4}\rangle$. Although nonzero transition elements exist between the state $|\psi_{1,4}\rangle$ and some two-excitation states such as  $|\psi_{2,2}\rangle$, these transitions are suppressed by the off-resonances $\sim \hbar J$ when $\epsilon\ll J$. The second pulse has opposite phases in the odd and the even sites with $(\epsilon_{1},\epsilon_{2},\epsilon_{3},\epsilon_{4})=\epsilon(1,-1,1,-1)$ and the pumping frequency $\omega_{d}=\omega_{0}-g-J$. Among the states $|\psi_{2,m}\rangle$ for $1\le m\le 6$, the only nonzero transition element induced by this pulse is $\langle\psi_{2,3}|H_{d}|\psi_{1,4}\rangle=\hbar \epsilon$. No coupling is induced by this pulse between the states $|\psi_{1,4}\rangle$ and $|\psi_0\rangle$, and hence there is no transition back to the ground state. Although nonzero transition elements exist between $|\psi_{1,4}\rangle$ and other $N=2$ states with higher energy such as  $|\psi_{2,7}\rangle$ and between $|\psi_{2,3}\rangle$ and some of the $N=3$ states, these transitions are again suppressed by the off-resonances $\sim \hbar g$. This pulse then induces a Rabi oscillation between $|\psi_{1,4}\rangle$ and $|\psi_{2,3}\rangle$. After being applied for a duration of $\pi/2\epsilon$,  this pulse generates the entangled polariton state $|\psi_{2,3}\rangle$.  

The entangled polaritons can be transformed into entangled photons by adiabatically switching the JC Lattice to the ``superfluid'' regime. To achieve this, we slowly increase the detuning to enter the dispersive regime with $\Delta\gg g$. During this process, the lower polariton modes  adiabatically evolve into photon Fock states and the nonlinear interaction is effectively turned off. The entangled state $|\psi_{2,3}\rangle$, as an eigenstate, adiabatically evolves into $(a_1^\dag a_3^\dag - a_2^\dag a_4^\dag)|\psi_0\rangle/\sqrt{2}$ which contains two spatially-entangled photons. The adiabatic condition for reaching this state requires that the increase of the qubit energy satisfies $d\omega_{z}/dt\ll 4g^2$, and hence the duration of the adiabatic process satisfies $\delta t \gg (\Delta/4g^2)$ \cite{EPAPS}. For our system, this condition can be readily met when $1/J < \delta t <1/\epsilon$. Meanwhile, this state can also be written in the momentum space. In the ``superfluid'' regime, the single-particle eigenmodes are
\begin{equation}
\left[\begin{array}{c}
c_{1}\\
c_{2}\\
c_{3}\\
c_{4}\end{array}\right]=\frac{1}{2}\left[\begin{array}{cccc}
1 & -1 & -1 & 1\\
-1 & 1 & -1 & 1\\
-1 & -1 & 1 & 1\\
1 & 1 & 1 & 1\end{array}\right]\left[\begin{array}{c}
a_{1}\\
a_{2}\\
a_{3}\\
a_{4}\end{array}\right]\label{eq:cm}
\end{equation}
with energies $\omega_0-2J$ for mode $c_2$, $\omega_0$ for modes $c_1$ and $c_3$, and $\omega_0+2J$ for mode $c_4$. It can be shown that the above entangled state can be written as $(c_1^\dag c_3^\dag - c_2^\dag c_4^\dag) \psi_0/\sqrt{2}$ and is hence entangled in the momentum space as well. By constructing coherent interface with outside circuits, the entangled photons can be distributed to outside modes and used in quantum information protocols.

The success of the proposed scheme relies on three criteria: 1. the energy scales are well separated with $\epsilon\ll J,g$; 2. the decoherence rates of the resonators $\kappa$ and qubits $\gamma_q$ satisfy $\kappa, \gamma_q \ll \epsilon$; 3. the fabrication errors of the system parameters can be compensated.  For superconducting resonators \cite{CQED_QO, MilburnKerrCQED}, strong coupling to qubits reaching $g/2\pi=100\,\textrm{MHz}$ has been demonstrated. The tunneling $J$ can be obtained as capacitive coupling between adjacent resonators. For a superconducting coplanar waveguide resonator, the voltage at position $x$ is 
\begin{equation}
V_{j}(x)=\sqrt{\frac{\hbar\omega_{0}}{C_r}}(a_{j}^{\dag}+a_{j})\cos(2\pi x/L)\label{eq:Voltage}
\end{equation}
with $L$ being the length of the quasi-one-dimensional resonator. With a coupling capacitance $C_{1}$ and resonator capacitance $C_{r}$, we derive the coupling as \cite{TianNJP2008} 
\begin{equation}
C_{1}V_{j}(L)V_{j+1}(0)=\hbar\omega_{0}\frac{C_{1}}{C_{r}}(a_{j}^{\dag}a_{j+1}+a_{j+1}^{\dag}a_{j})\label{eq:coupling}
\end{equation}
which yields $\hbar J=\hbar\omega_{0}(C_{1}/C_{r})$. The tunneling can be adjusted in a large range by varying $C_{1}$.  For example, $\omega_{0}/2\pi=10\,\mathrm{GHz}$, $C_{1}=10\,\text{fF}$, and $C_{0}=2\,\text{pF}$, we have $J/2\pi=50\,\mathrm{MHz}$. 

For finite driving amplitude $\epsilon$, the off-resonant transitions occur with the probability $\sim |\hbar\epsilon/\delta E|^{2}$ depending on the off-resonance $\delta E$. For example, as is shown in Fig.~\ref{fig2}, the second pulse in this scheme induces transition between $|\psi_{1,4}\rangle$ and $|\psi_{2,7}\rangle$ with the  probability $\sim|\epsilon/0.6g|^{2}$.  Decreasing the driving amplitude can decrease the off-resonant transitions. However, it also means longer pulse duration and larger influence from the decoherence of the resonators and qubits. The damping rate of the resonators can be as low as $\kappa/2\pi\sim 10\,\textrm{kHz}$ given a quality factor $Q=10^{6}$ from recent experiments. The decoherence rate of the qubits can be less than $\gamma_{q}/2\pi\sim100\,\textrm{kHz}$.  At $\Delta=0$, the decoherence rate of the polariton modes is $\gamma_p\sim (\kappa+\gamma_{q})/2$ which can be less than $50\,\textrm{kHz}$. In the dispersive regime, the decoherence rate is mostly determined by the damping rate of the resonator. A driving amplitude of $\epsilon/2\pi=5\,\text{MHz}$ can hence be adopted to achieve high fidelity in the generation of entangled states.

\begin{figure}
\includegraphics[clip,width=8cm]{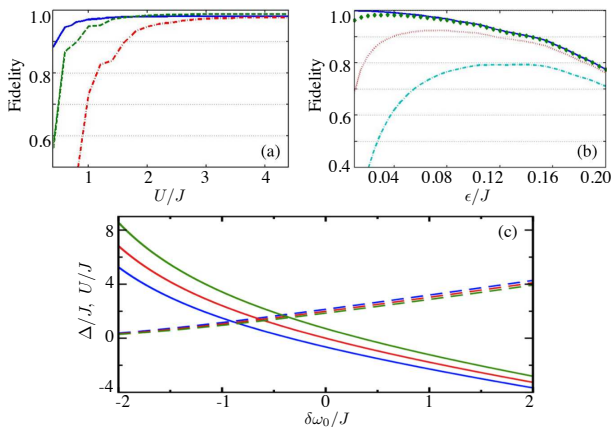} 
\caption{(a) Fidelity versus $U/J$ at $\epsilon/J=0.02,\,0.04,\,0.1$ from top to bottom and $\gamma_{p}/J=2\times 10^{-4}$. (b) Fidelity versus $\epsilon/J$ at $\gamma_{p}/J=0,\,2\times 10^{-4},\,2\times 10^{-3},\,0.01$ from top to bottom at $U/J=2$. (c) Solid (dashed) curves: detuning (effective $U/J$) versus frequency errors $\delta\omega_0$ at coupling errors $\delta g=0.1g,\,0,\,-0.1g$ from top to bottom (from bottom to top). And $J/2\pi=50\,\textrm{MHz}$. }
\label{fig3}
\end{figure}

To test this analysis, we numerically simulate this scheme using a master equation approach and calculate the fidelity of the final density matrix: $F=\langle \psi_{2,3}|\rho_f|\psi_{2,3}\rangle$ with the target state $|\psi_{2,3}\rangle$. For simplification, we use the energy spectrum in Fig.~\ref{fig2}(b) in our simulation, which can be treated as a Bose-Hubbard Model for the lower-polariton modes $L_{j,1}$ with damping rate $\gamma_p$. The effective self-interaction is defined as $U= (E_{2-} - 2E_{1-}) /\hbar$ in terms of the polariton energies \cite{CQED} and $U=(2-\sqrt{2}) g$ at $\Delta=0$ when the entangled photons are generated. The other energy levels in Fig.~\ref{fig2}(a) are omitted. This model, though simplified, captures the dominant effects that reduce the fidelity: the off-resonant transitions and damping. The damping terms in the master equation are the standard Lindblad form for modes $L_{j,1}$ \cite{QOBook}. In Fig.~\ref{fig3}(a), the fidelity increases rapidly with $U$ in the regime $U<J$. For $U>J$, the fidelity stops increasing with $U$ and reaches a saturation value that does not depend strongly on the driving amplitude. This agrees with our previous analysis that the fidelity is limited by the off-resonance $\delta E\sim \min(\hbar J, \,\hbar U)$ when $\gamma_p\ll \epsilon$. In Fig.~\ref{fig3}(b), we find that the fidelity increases with the driving amplitude $\epsilon$ at first, but soon reaches a maximum and starts to decrease as $\epsilon$ further increases.  The initial increase of the fidelity is due to the shortening of the pulse durations and hence the lessening of the damping as $\epsilon$ increases. When $\epsilon$ further increases pass the maximum, the off-resonant transitions gain a larger probability which eventually makes the fidelity decrease. Our result shows that the fidelity can exceed $0.9$ at $\gamma_p\sim 100\,\textrm{kHz}$ ($\gamma_p/J\sim2\times 10^{-3}$).

The parameters of the superconducting devices can spread by a few percent due to fabrication errors. The magnitude of the deviations in the resonator and qubit frequencies can be comparable with the coupling and the tunneling constants. Mostly coming from the in-accuracies in the Josephson energy, the deviations in qubit frequency can be avoided because the qubits are made of SQUID loops instead of single junctions. The qubit frequency as well as the detuning $\Delta$ is hence tunable. Below we will show that by adjusting the detuning, the effect of the deviations in the resonator frequency and the deviations in the coupling constant on the energy of the $L_{j,1}^{\dag}|\psi_{0}\rangle$ modes can also be fully compensated. The energy of the $L_{j,1}^{\dag}|\psi_{0}\rangle$ modes is $E_{1-}/\hbar = \omega_0 + \Delta/2 - \sqrt{\Delta^2/4+g^2}$ \cite{CQED}. Assuming the deviations $\delta\omega_0$ in the resonator frequency and $\delta g$ in coupling strength, we find that the detuning needs to be
\begin{equation}
\Delta=[(\delta g + g)^2-(\delta\omega_0+ g)^2]/(\delta\omega_0+g)\label{eq:Delta}
\end{equation}
in order to recover the energy $E_{1-}/\hbar=\omega_0-g$ for the lower polariton mode at $\Delta=0$. This detuning is plotted in Fig.~\ref{fig3}(c) for various $\delta\omega_0$ and $\delta g$. Note that the energy of the $T_{j,1}^{\dag}|\psi_{0}\rangle$  modes can not be compensated simultaneously. As a result, the interaction $U$ after the compensation varies with $\delta\omega_0$ and $\delta g$ and can be decreased for a negative $\delta\omega_0$. To ensure a large $U$ to protect the fidelity of the pumping scheme, we can choose the lowest resonator frequency in the array as the base frequency $\omega_0$ so that $\delta\omega_0>0$ for all other resonators. 

In summary, we presented a scheme to generate entangle photons in coupled superconducting resonators. Our scheme explores the Kerr nonlinearity in a JC Lattice, the circuit symmetry, and the spectroscopic techniques to pump entangled photon pairs with high fidelity. We numerically studied the robustness of this approach in the presence of decoherence. The idea can also be easily generalized to semiconductor system. When applied to multiple resonators,  this idea can shed new light in the quantum engineering of novel states of microwave photons \cite{GHZ}.

\emph{Acknowledgements.} This work is supported by the National Science Foundation under Grant No. NSF-CCF-0916303, NSF-DMR-0956064, and by UC Merced funds.

\end{document}